\documentclass[letterpaper,11pt]{emulateapj}
\usepackage{graphicx}
\usepackage{color}
\usepackage{dashrule}
\usepackage{multirow,bigstrut,bigdelim}
\usepackage{epsfig}

\usepackage{graphics}
\usepackage{graphicx}
\begin{document}
\title{Magnetically and Baryonically Dominated Photospheric\\ Gamma-Ray Burst Model
Fits to Fermi LAT Observations}
\author{P\'eter Veres\altaffilmark{1,*},  Bin-Bin Zhang\altaffilmark{2}, P\'eter \Mesz\altaffilmark{1}}
\date{\today} 
\altaffiltext{1}{Department of Astronomy and Astrophysics, Department of
Physics, and Center for Particle and Gravitational Astrophysics,
Pennsylvania State University, 525 Davey Lab, University Park, PA 16802, USA}
\altaffiltext{2}{Department of Astronomy and Astrophysics,
Pennsylvania State University, 525 Davey Lab, University Park, PA 16802, USA}
\altaffiltext{*}{Email: veresp@psu.edu}

\date{\today} 
\def\ve{\varepsilon}
\def\gbm{{\it GBM }}
\def\lat{{\it LAT }}
\def\fermi{{\it Fermi }}
\newcommand{\Mesz}{{M\'esz\'aros}}
\def\mathnew{\mathsurround=0pt}
\def\simov#1#2{\lower .5pt\vbox{\baselineskip0pt \lineskip-.5pt
       \ialign{$\mathnew#1\hfil##\hfil$\crcr#2\crcr\sim\crcr}}}
\def\simg{\mathrel{\mathpalette\simov >}}
\def\siml{\mathrel{\mathpalette\simov <}}
\def\beq{\begin{equation}}
\def\enq{\end{equation}}
\def\bea{\begin{eqnarray}}
\def\ena{\end{eqnarray}}
\def\bitm{\bibitem}
\def\msun{M_\odot}
\def\L54{L_{54}}
\def\E55{E_{55}}
\def\et3{\eta_3}
\def\th1{\theta_{-1}}
\def\r07{r_{0,7}}
\def\x05{x_{0.5}}
\def\et600{\eta_{600}}
\def\et3{\eta_3}
\def\rph{r_{ph}}
\def\vareps{\varepsilon}
\def\fflunit{\hbox{~erg cm}^{-2}~\hbox{s}^{-1}}
\def\eps{\epsilon}
\def\ve{\varepsilon}
\def\muh{\hat{\mu}}
\def\cm{\hbox{~cm}}
\def\s{\hbox{~s}}
\def\gev{\hbox{~GeV}}
\def\GeV{\hbox{~GeV}}
\def\MeV{\hbox{~MeV}}
\def\kev{\hbox{~keV}}
\def\keV{\hbox{~keV}}
\def\eV{\hbox{~eV}}
\def\erg{\hbox{~erg}}
\def\s{{\hbox{~s}}
\def\cm2{\hbox{~cm}^2}}
\def\para{\parallel}
\defcitealias{Veres+12magnetic}{VM12}

\begin{abstract}
We consider gamma-ray burst models where the radiation is dominated by a
photospheric region providing the MeV Band spectrum, and an external shock
region responsible for the GeV radiation via inverse Compton scattering.  We
parametrize the initial dynamics through an acceleration law $\Gamma\propto
r^\mu$, with $\mu$ between 1/3 and 1 to represent the range between an extreme
magnetically dominated and a baryonically dominated regime{, depending also
on the magnetic field configuration}.  We compare these models to several
bright \fermi \lat bursts, and show that both the time integrated and the time
resolved spectra, where available, can be well described by these models. We
discuss the parameters which result from these fits, and discuss the relative
merits and shortcomings of the two models.  \end{abstract}

\section{Introduction}
\label{sec:intro}

Observations in the GeV range of gamma-ray bursts (GRBs) by the Fermi LAT
\citep{Atwood+09LAT} instrument have uncovered new peculiarities of these
objects, which were unknown previously. Among the  minority of bursts which
show significant emission above 100 MeV \citep{Kocevski+12LATpaucity}, we
concentrate here on a handful of bursts with substantial detected GeV emission
to construct models aimed at interpreting the radiation and the basic
parameters of the bursts.

The bursts with observed GeV emission have rather diverse spectra. So far,
because of relatively low number of such GRBs, we restrict ourselves only to
bursts which have a large enough data coverage to make detailed model fits
meaningful.  In general terms, the GeV emission in some bursts can be described
through an extrapolation of the lower energy spectrum, without strong evidence
for departures from this, while in other bursts an extra power law component
emerges at GeV energies, in addition to the spectral component describing the
MeV radiation.  An important new feature present in both short and long bursts
is the presence of a delay between the MeV and the GeV radiation. Furthermore,
some bursts also exhibit strong spectral evolution.

The origin of the GeV component has been attributed to leptonic mechanisms in
the external forward shock, e.g. in the radiative regime in a pair enriched
interstellar medium \citep{Ghisellini+09-gevrad} or in the adiabatic regime
\citep{Kumar+09gevforsho}, or alternatively to hadronic cascades
\citep[e.g.][]{Asano+09-090510}, among others.  In the external forward shock,
inverse Compton scattering and Klein-Nishina effects were considered by
\citet{He+12grbnu}, and it has been argued that at least the first few seconds
of the early high energy emission cannot be attributed to the forward shock
{\citep{He+11-090510,Gao+09HE}}.  Thus, it is important to consider both the prompt
(largely MeV) emission and the GeV emission jointly, as they interact with each
other. Two recent developments which are relevant for the prompt emission are
the increasing consideration of magnetically dominated models for the GRB
outflow, and of photospheric models for the formation of the prompt radiation
spectrum \citep[e.g.][]{Meszaros+12raa}.

Previously, we have shown  that some of the  brightest \lat detected bursts can
be modelled by a magnetic two component model \citep{Veres+12magnetic}
\citepalias[][hereafter]{Veres+12magnetic}.  We showed that within this model,
without going into detailed model fits, it is possible to incorporate the
seemingly different behaviors of the high-energy spectrum (which is either a
single Band function or a Band function with an extra high energy component)
into a single framework.  We also addressed the non-detection problem of bursts
by LAT, namely some of the GRBs should have been detected by LAT if the
spectrum continues smoothly from the MeV regime, or generally there is lack of
LAT detected bursts compared to initial estimates, based on the simple
extrapolation of the MeV emission. We have also presented a generalized
formalism for the dynamics of GRBs valid for a generic acceleration law
$\Gamma\propto r^\mu$ with $1/3\le\mu\le1$ \citep{Veres+12-110721a}, where
$\mu$ can be understood as a parameter accounting for the bulk Lorentz factor
$\Gamma(r)$ averaged across the effective cross section of the jet at that
radius.

Here we present detailed numerical fits of the MeV and GeV data to specific
physical models of GRBs based a photospheric origin of the prompt emission in
two dynamic regimes, i.e. the extreme magnetically dominated regime and the
usual baryonically dominated regime, with inclusion of the inverse scattering
effects (both self and external) incurred in the external forward and reverse
shocks.  This is one of the few instances of fitting GRB spectra with
physically detailed  models (for other examples see \citet{Ryde+10phot090902,
Burgess+11synch}). We construct the models based on the generalized dynamics
described above, and convolve them with the response function of the \fermi
detectors, showing that they give a good statistical description of the data.

In this paper, we analyze four representative bursts, which capture the
diversity observed in LAT: GRBs 080916C, 090510, 090902B and 090926A. We model
their spectrum with the two component magnetic ($\mu=1/3$) model and
demonstrate that it is a viable model to interpret the spectra from
approximately 10 keV to 10 GeV. We also show that the baryonic model ($\mu=1$)
gives similarly good fits to the data, as quantified by specific goodness of
fit criteria.

We present our basic  model in \S 2. The Fermi observational data for the four
GRBs and the fitting method are described in \S 3. The fitting results are
presented in \S 4. We summarize the results and compare the baryonic with the
magnetic model, with a discussion on the physical implications, in \S 5.

\section{Photospheric model with arbitrary acceleration parameter}
\label{sec:model}

Our basic  model consists of a dissipative photosphere, whose location depends
on whether the dynamics there is kinetic energy dominated or magnetically
dominated, and an external shock region, where reverse shocks may be present as
well as a forward blast wave. The bulk of the prompt MeV radiation arises from
the neighborhood of the photosphere in the form of a nonthermal
component\footnote{Depending on model details, the spectrum may start forming
below the photosphere, and further contributions can arise from outside of it
\citep{Peer+06phot,Lazzati+10phot,Beloborodov10phot, Giannios11peak}; for our purposes here, we can assume
that the bulk of the radiation arises at the photosphere.}, and depending on
the model, a separate, subdominant thermal component can also be evident.  At
the deceleration radius, shock accelerated electrons upscatter the prompt
photons to produce a component peaking in the GeV range.

In the present paper we generalize the model presented in \citetalias{Veres+12magnetic}
to cover a wider range of dynamic behavior stretching from the extreme kinetic energy
dominated to the extreme magnetic dominated cases.  The acceleration of the ejecta is
described by the expression (\ref{eq:accel}) for the bulk Lorentz factor
\begin{equation}
{\Gamma(r)}\propto \left\{
\begin{array}{lll}
 r^{\mu}		&	{\rm if}	& r<r_{\rm sat}\\
 {\rm const.}	&	{\rm if}	& r_{\rm sat}<r<r_{\rm dec}.
\end{array}
\right.
\label{eq:accel}
\end{equation}
Here the exponent $\mu$ is taken as $\mu=1/3$ when the dynamics is
governed by dissipation of the energy stored in magnetic fields in the presence of
a sub-dominant inertia of the baryonic component \citep[e.g.][]{Drenkhahn02,
Meszaros+11gevmag}, while if the baryonic component dominates, we have	$\mu=1$.
Intermediate values of	$\mu$ can arise in magnetic outflows with certain symmetries{, and the exact value of $\mu$ also depends on the magnetic field configuration}
\citep[e.g.][]{Tchekhovskoy+10grb,McKinney+11switch}.   $r_{\rm sat}$
and $r_{\rm dec}$ are the saturation (coasting) and the deceleration radii, and
with this parametrization the radius where the ejecta reaches the coasting Lorentz
factor is $r_{\rm sat}= r_0 \eta^{\frac{1}{\mu}}$. For simplicity we consider the
transition from the acceleration phase to coasting as a broken power law function,
although this transition occurs more gradually \citep[e.g.][]{kobayashi99}.

In what follows is it is useful to give the dependence of the physical parameters
in terms of a general $\mu$. The main difference between the models is whether
the photosphere occurs below or above the saturation radius. This is governed by
the relation of $\eta$ and a quantity $\eta_T$ given by eq.(\ref{eq:etat}).
Here $\eta_T$ is a limiting value for the Lorentz factor, which separates the two
cases of $\eta>\eta_T$ (photosphere below saturation) and $\eta<\eta_T$ (photosphere
above saturation, where $\eta_T$ is a function of the dynamics, depending on $\mu$.
This limiting Lorentz factor is obtained from setting $r_{\rm sat}=r_{\rm ph}$ for
a general $\mu$, generalizing the related value introduced in \citet{Meszaros+93gasdyn},
\begin{equation}
\eta_{T}=\left(\frac{ L\sigma_T}{8\pi m_p c^3 r_0}\right)^{{\mu}/{(1+3\mu)}}.
\label{eq:etat}
\end{equation}
The photosphere occurs in the acceleration phase if $\eta>\eta_T$, typical for
a magnetically dominated ($\mu=1/3$) case, where $\eta_T\simeq 150~L_{53}^{1/6}
r_{0,7}^{-1/6}$.  The photosphere is in the coasting phase for $\eta<\eta_T$ which it
is typical for baryonic cases ($\mu=1$) for which $\eta_T\simeq 1900~L_{53}^{1/4}
r_{0,7}^{-1/4}$.  The two radii are equal for $\mu\approx0.6$ if other nominal
parameters are left unchanged.	The photosphere will occur at
\begin{eqnarray}
\frac{r_{\rm ph}}{r_0}&=&\left(\frac{ L\sigma_T}{8\pi m_p
c^3r_0}\right)\frac{1}{\eta \Gamma_{\rm ph}^2} =\\
&=&\left\{
\begin{array}{ll}
 \eta_T^{1/\mu}(\eta_T/\eta)^{1/(1+2\mu)}	&	{\rm if~ }	 \eta>\eta_T \\
 \eta_T^{1/\mu}(\eta_T/\eta)^3			&	{\rm if~ }	 \eta<\eta_T
\end{array}
\right.
\end{eqnarray}
noting that $\Gamma_{\rm ph}$ carries an $r_{\rm ph}$ dependence for
$\eta>\eta_T$. The Lorentz factor at the photosphere is $\Gamma_{\rm
ph}=(r_{\rm ph}/r_0)^{\mu}\propto L^{\mu/(2\mu+1)} \eta^{-\mu/(2\mu+1)}
r_0^{-\mu/(2\mu+1)} $ for $\eta>\eta_T$ and $\Gamma_{\rm ph}\approx\eta$ in
other cases.
The shell has initially a constant width, $\Delta r = c t_w\approx 3 \times
10^{11}  t_{1} \cm$, where $t_w$ is the time during which the central engine is
active. At a later stage the shell starts to spread radially when the
when the relative width $\delta r/r\sim \Delta v/c\sim 1/2\Gamma^2$ becomes of the
same order as the initial shell width $\Delta r /r$.  Depending on the regime, this
can happen at $r_{\rm spread}=
 \left(2 c t_w/r_0^{2\mu}\right)^{1/1-2\mu}	{\rm if } ~r<r_{\rm sat} $ and $
 2 \eta^2 c t_w 	 {\rm if } ~r>r_{\rm sat}$.
For most of the parameters used here the spreading occurs well beyond the
saturation radius, $r_{\rm spread}=6\times 10^{17} \eta_{600}^2 t_1\cm $, and
it can be neglected close to the photosphere.
The scaling of the baryon number density in the comoving frame is
\begin{equation}
n'_b(r)=L/4\pi r^2 m_p c^3 \Gamma(r) \eta.
\label{eq:nb}
\end{equation}
This results in a dependence:
$n'_b(r)\propto V'^{-1}(r)\propto r^{-2-\mu} $ if $ ~r<r_{\rm sat}$ or	$
r^{-2}$ if $ r>r_{\rm sat}$.

{Prompt spectrum:} Observationally the prompt emission spectrum is described by
a Band function \citep{Band+93}, and a spectrum with broadly similar attributes is
shown to arise in photospheres \citep{thompson94,Beloborodov10phot,Vurm+11phot,Giannios11peak,
Peer+12thermal,Beloborodov12peak}.  The location of the peak of the spectrum may be
due to effects related to electron heating by nuclear collisions, reconnection, or
semirelativistic shocks, all of which give the peak energy at the right order of
magnitude. A detailed radiative transfer would be model-dependent and too
cumbersome for performing numerical fits to the data. For the purposes of this study
we choose a simple prescription where the peak energy is determined by the synchrotron
peak from mildly relativistic shocks around the photosphere\footnote{Shocks are
expected as result of magnetic reconnection, as well as possibly from outflow
Lorentz factor fluctuations, the particle acceleration time being of the same order
for reconnection and Fermi acceleration \cite{Meszaros+11gevmag}, and the spectral peaks
are in both cases in the MeV range, as for baryonic thermal photosphere models.}.

It should be noted however, that the above is just a simple heuristic
prescription for the peak, the shape being taken to be the Band shape.  The
Band spectra and peaks deduced in the previously mentioned numerical
simulations arise from distortions of the original black-body component formed
around the launching radius through Compton scattering and synchrotron
radiation, which escapes at the photosphere. In these cases no additional
black-body component is expected below the peak. In our model considered here
the peak is ascribed to synchrotron radiation from electrons accelerated near
the photosphere, which introduces a dependence on the magnetic field; and the
advected blackbody component is also released, but at energies typically below
the peak, for magnetically dominated dynamics. There are possible shortfalls
for a pure synchrotron interpretation, such as the low energy spectral slope
observations which show inconsistencies with the synchrotron model when electrons are
in the fast cooling regime \citep[e.g.][]{Preece+98syn}. While recent analyses
\citep{Sakamoto+11batcat} show that the fraction of bursts showing problems is smaller
then previously thought, there is still a non-negligible fraction of bursts
where other mechanisms may need to be invoked. This could be the Compton
upscattering mentioned above, or modifications of the simple synchrotron
mechanism, e.g. such as discussed by \citet{Medvedev06jitter,Daigne+11syn}. A first
principles discussion of such cases for our fitting exercise undertaken here on
individual bursts would be too extensive, and for this reason we have adopted
the simple prescription mentioned above.

The peak of the prompt spectrum is then given by $\ve_b=({3q_e B'_{\rm ph}}/
{4 \pi m_e c}) \gamma_{\rm e,ph}^2 [{\Gamma_{\rm ph}}/({1+z})]$, where $B'\propto
{n'}_b^{1/2} \Gamma_r$ {(taking $n'_b$ from Eq. \ref{eq:nb})} and $\gamma_{e,ph}\propto \Gamma_r$
\begin{eqnarray}
\ve_b \propto \left\{
\begin{array}{ll}
  L^{\frac{3\mu-1}{4\mu+2}} \eta^{-\frac{3\mu-1}{4\mu+2}}
  r_0^{\frac{-5\mu}{4\mu+2}} \Gamma_r^3/(1+z)		&	{\rm if~ }  \eta>\eta_T\\
  L^{-1/2} \eta^{3} 
    \Gamma_r^3/(1+z)		&	{\rm if~ }  \eta<\eta_T.
\end{array}
\right.
\label{eq:ebreak}
\end{eqnarray}
{Here we have assumed an efficiency of 0.5 \citep{Giannios11peak}.}
The two cases imply widely different values for the same set of parameters.
This is mainly caused by the strong dependence on $\eta$  when the
photosphere arises in the coasting phase ($\eta<\eta_T$).  Note furthermore,
that the photosphere loses any memory of the acceleration in the same case and
no $\mu$ dependence is present. Also in this case, the peak energy is
independent of the launching radius.  We take the low and high energy photon
number indices from observations ($\alpha\approx 1$, $\beta\approx2.2$), and in
the fitting sections we directly fit for these.

{Thermal component: } Even though the main emission component is nonthermal, one
expects a thermal component to arise from the photosphere. There are hints of
this separate thermal component emerging below the Band peak
\citep{Page+11Xthermal, Guiriec+10-100724, mcglynn+12phot}. The comoving
temperature goes as ${T'}\propto r^{-\left(\frac{\mu+2}{3}\right)}	$ if $
r<r_{\rm sat} $ and $ r^{-2/3}	$ if $	 r_{\rm sat}<r<r_{\rm spread}.$
This results in a dependence of the observed temperature as:
\begin{eqnarray}
T_{\rm obs}(r_{\rm ph}) \propto \left\{
\begin{array}{ll}
  L^{\frac{14\mu-5}{12(2\mu+1)}} \eta^{\frac{2-2\mu}{6\mu+3}}
  r_0^{-\frac{10\mu-1}{6(2\mu+1)}} /(1+z)	& {\rm if~ } \eta>\eta_T \\
  L^{-5/12} \eta^{8/3} r_0^{1/6} /(1+z) 	&	{\rm if~ } \eta<\eta_T .
\end{array}
\right.
\label{eq:temp}
\end{eqnarray}

{External shock-related components:}
The ejecta starts to decelerate when it has plowed up approximately $1/\eta$
times its own mass from the interstellar matter. This occurs at the
deceleration radius $ r_{\rm dec}= 4.8 \times 10^{16}	L_{t,53}^{1/3}
(1-\zeta_r)^{1/3} t_{1.3}^{1/3} n_0^{-1/3} \eta_{600}^{-2/3} \cm$. Here
$\zeta_r$ is the fraction of the luminosity in radiation, which for simplicity
is fixed to $0.5$ throughout the rest of this article.	At the	deceleration
radius a forward shock develops (and possibly a reverse shock as well). The
prompt emission is up-scattered on the shock-accelerated electrons, giving
rise, to a $\gtrsim 1 \gev$ peaked radiation. This component dominates at
sufficiently high energies, but radiation from synchrotron self Compton (SSC)
from the forward shock (FS) may also be relevant here.	The deceleration radius
is larger than the photospheric and the saturation radius for every realistic
set of parameters.

\section{Data and fitting method}
\label{sec:datafit}
\subsection{Data}
\label{sec:data}

Currently there are at least 40  LAT-detected GRBs at energies $\gtrsim 100 $
MeV \citep{Omodei+11fermigrb,Zheng+12extraLAT}.  Out of these, there are only a
handful of bursts with	enough data at $\gev$ energies to derive meaningful
constraints on the model parameters \citep{Zhang+11-latgrb}. In this paper we
choose four of such bright LAT GRBs, namely, GRB 080916C, 090510, 090902B and
090926A as our sample.

We processed the Fermi GBM and LAT data using the same method as presented in
\citet{Zhang+11-latgrb}.  For each burst, we choose the 1-2 brightest GBM
spectra and 1 or 2 BGO spectra together with LAT spectrum to compare with our
model.  We have fixed the BGO normalization factor, accounting for differences
between detectors' effective area, to the usual value of $0.9$
\citep{Abdo+09-090902B}.

By varying the free parameters,
we get the best models which can reproduce the observed count spectrum.  For the total
duration of the bursts we managed to fit GRBs 080916C and 090510 with our model.
For GRBs 090926A and 090902B, our model did not give a satisfactory fit when
considering the entire duration of the burst, but time resolved spectra were well
fitted by our model. We put this on the account of the strong spectral evolution
present in these two GRBs as published in their initial discovery papers.

\subsection{Fitting method}
\label{sec:fitmet}
We have taken the theoretical model described above with 9 free parameters.  We
fit the data using the simplex method implemented in the {\it amoeba} routine
\citep[see ][]{nr} modified, in some of the cases with simulated annealing
(Erik Rosolowsky, private communication). This addition makes the method less
sensitive to local minima.  The parameters describe the spectrum  of a
photospheric model (either magnetic or baryonic).

The procedure is to convolve the model with the detectors' response functions
and compare this result with the observed counts.  For this purpose, we have
written our own IDL codes to handle both the observed count spectrum, the
instrumental response matrix (DRM) and read in the multi-parameter theoretical
model to carry out the fitting. 

Because the number of fitting parameters is large, we introduce some additional
constraints, in accordance with the model. We require the deceleration time to
be of order of few seconds, which corresponds to the delay between the MeV and
GeV components. This is achieved by introducing a penalty in the goodness of
fit for $t_{\rm dec} \gtrsim $ few $s$.  We vary one by one each of the
parameters around the best fit to calculate their errors.  When the $\chi^2$
statistic increases by the amount characteristic for the one-sigma limit for
given degrees of freedom, we fix the range and thus the errors for that
parameter.

\begin{table}
\begin{center}
  \begin{tabular}{lcccc} \hline
  & 080916C(m) & bar &  090510(m) & bar    \\
\hline \hline
 $\epsilon_e$/0.01	&   $3.1^{+0.6}_{-0.7}$ 	&$2.0^{{+}1.7}_{{-}1.0}$	& $0.50^{+0.50}_{-0.25}$			& $8.3^{+8.3}_{-4.2}$		 \\
 $\eta$ 			&  $1277^{+111}_{-150}$ 	&$312^{+11}_{-8}$		& $318^{+77}_{-159}$			& $235^{+18}_{-14}$		 \\
 $ \epsilon_{\rm B,FS}/10^{-3}$
					& $56^{+54}_{-32}$			&$0.022^{+0.27}_{-0.017}$	& $4.4*$							& $0.49 *$			\\
 $ L_t [10^{53}\frac{\rm erg}{\rm s}]$
					&  $76\pm0.6$				&$20^{+1}_{-1.1}$		& $0.57{\pm 0.11}$				& $0.60^{+0.08}_{-0.09}$	  \\
 $ \alpha $		& $0.47^{+0.13}_{-0.14}$	&$0.95^{+0.06}_{-0.08}$ & $0.77^{+0.15}_{-0.20}$		& $0.55^{+0.20}_{-0.31}$	  \\
 $ \beta $			& $1.92^{+0.01}_{-0.01}$	&$2.11\pm0.01$			& $4.49^{+0.02}_{-1.04}$		& $4.42^{+0.07}_{-1.03}$	   \\
 $ n_{\rm e} [{\rm cm}^{-3}]  $
					&   $16^{+11}_{-9} $		&$64^*$ 				& $4.1\times 10^3 *$				& $1.1\times 10^5 *$	\\
 $ p $				&  $2.87^{+0.12}_{-0.08}$	&$2.1^{+2.9}_{-0.05}$	& $4.74^{+0.26}_{-2.73}$		& $2.02^{+2.98}_{-0.01}$	    \\
 $ \Gamma_r $		&   $1.32 \pm 0.05$		&$1.0\pm 0.03 $ 		& $2.24^{+0.16}_{-0.13}$		& $1.00^{+0.07}_{-0.0}$ 	    \\
 $\chi^2$/dof		& $ 113.6 /  98$			&$129.7 / 98 $			& $107.2 /  97$ 				& $ 107.1 /  97$		    \\
\hline
\end{tabular}
\caption{Model parameters for the time-integrated fits with a magnetically and a
baryonically dominated model for two bursts. Stars mark parameters for which we could not determine the errors. }
\end{center}
\end{table}

\noindent
{\it Parameters constrained:}
In a situation when the spectrum is described by either an extremal magnetic
($\mu=1/3$) or baryonic ($\mu=1$) model,  the varying parameters we set out to
constrain are: the break energy ($\ve_b$, see Eq. \ref{eq:ebreak}) of the prompt
emission, the low- and high energy indices ($\alpha$ and $\beta$) of a Band
function ($N_{\ve}(\ve) \propto \ve^{-\alpha} e^{-\ve(2-\alpha)/\ve_{b}}$ for
$\ve < \ve_b$ and $N_{\ve}(\ve) \propto \ve^{-\beta}$ if $\ve > \ve_b$), the
relative Lorentz factor of the reconnecting regions in the photosphere
$\Gamma_r(\gtrsim 1)$, the total luminosity $L_t$, interstellar density
($n_e$), index of the electron power-law distribution ($p$, where $N_e(\gamma)
\propto \gamma^{-p}$) and the fraction of the energy in the magnetic field and
electrons at the deceleration radius ($\epsilon_{\rm B,FS}$ and $\epsilon_e$).
We do not to include variations in e.g. magnetic field in FS and RS because it
would not be possible to constrain them and we assume that $\epsilon_{\rm
B,FS}=\epsilon_{\rm B,RS}$. The components included in this work are: prompt
Band component (PR), blackbody component (BB), forward- and reverse shock
synchrotron (FS, RS), forward and reverse shock synchrotron self Compton
(FS-SSC, RS-SSC), external inverse Compton of the prompt photons on the forward
and the reverse shock (FSEIC, RSEIC).

The prompt emission is represented as a Band function, the forward and reverse
shock  synchrotron and inverse Compton components as multiple broken power
laws.  We have applied a phenomenological smoothing at the breaks of the type
$( (x/x_{\rm break})^{i_1 b}+(x/x_{\rm break})^{i_2 b})^{1/b}$.  Here $b$ is an
arbitrary constant describing the smoothness of the break (we set $b=-4$),
$i_{1,2}$ are the power law indices and the break occurs at $x_{\rm break}$.
The prompt emission may also have a high energy spectral cut-off, at an energy
depending on the acceleration mechanism at the photosphere, and whether or not
pair creation is present.  Here we have taken a phenomenological cut-off value
at approximately $\Gamma_{\rm ph} m_e c^2 /(1+z)$ (as also in
\citet{thompson94}).  Both without pair formation and for pair formation
estimated from the photons above threshold estimated from an extrapolation of
the observed spectral slopes in the bursts considered here, a cutoff close to
the above value is obtained. The energy coverage is generally sparse at a
putative prompt cutoff energy in the range of  tens of MeV, making it hard to
discern whether there is a pair cutoff or a cutoff due to the particle
acceleration mechanism.  For simplicity, here we use the model without pair
creation for the purposes of the numerical calculation.  This cutoff can
constrain the photospheric Lorentz factor ($\Gamma_{\rm ph}$).	Concerning the
GeV component, we identify the high energy cutoff (and peak) with the
Klein-Nishina cutoff.

The strength of the GeV component is significantly influenced  by the amplitude
of the prompt component (responsible for the intensity of seed photons for the
EIC scattering), $\epsilon_e$ and $n_{\rm ext}$, while	$\epsilon_{\rm B,FS}$
plays a role in some of the cases.  The shape of the GeV emission is defined by
a combination of $\alpha$ and $p$, which makes it possible to constrain these
parameters. The peak energy of the GeV emission depends on $\eta$, $\epsilon_e$
and $p$. The coasting Lorentz factor, $\eta$ can be constrained by the prompt
emission cutoff at tens of MeVs.

The model in general has difficulty in constraining $\epsilon_{\rm B,FS}$, as
this parameter affects only the  low-energy ($\simeq 10 \keV$) behavior of the
spectrum  through the FS synchrotron (and to a lesser extent by RS-SSC
emission). As a consequence this energy range can be used to constrain the
parameter, but in many cases, however, there are no signs of an extra component
emerging at low energies.  We note that, considering the small number of
high-energy photons, the fact that many of the combined GBM-LAT spectra are
consistent with a single Band function does not necessarily exclude additional
components.  The two components (photospheric and external shock-related) can
align to form a continuous component without  excessive fine tuning.  This
requirement is relaxed when we need to fit bursts  with additional components.

\section{Individual GRBs}
\label{sec:individual}

\subsection{GRB 090510 fit for time integrated spectrum}
\label{sec:090510}
\begin{figure}[htbp]
\begin{center}
\includegraphics[width=.7\columnwidth]{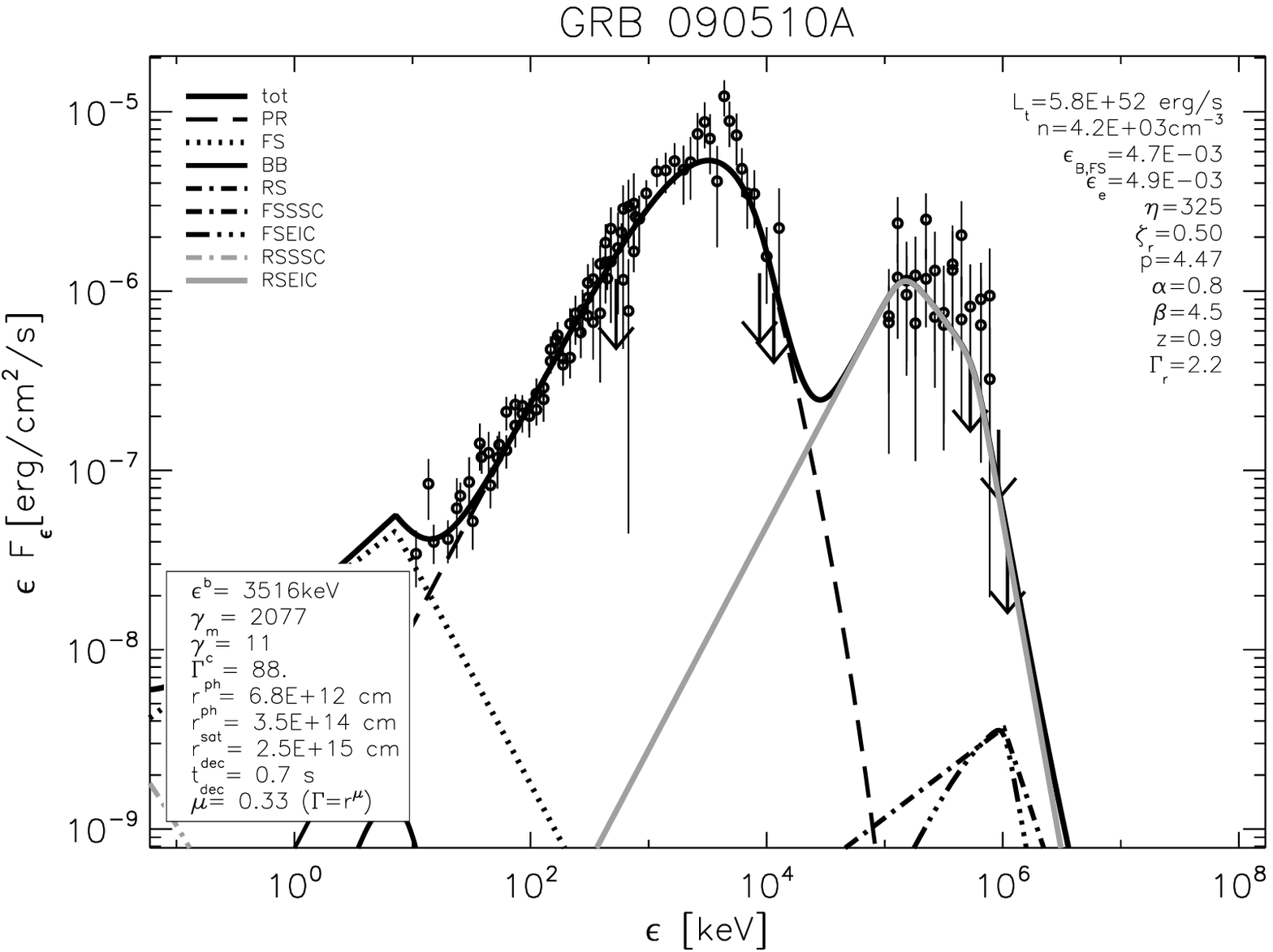}\\
\includegraphics[width=.7\columnwidth]{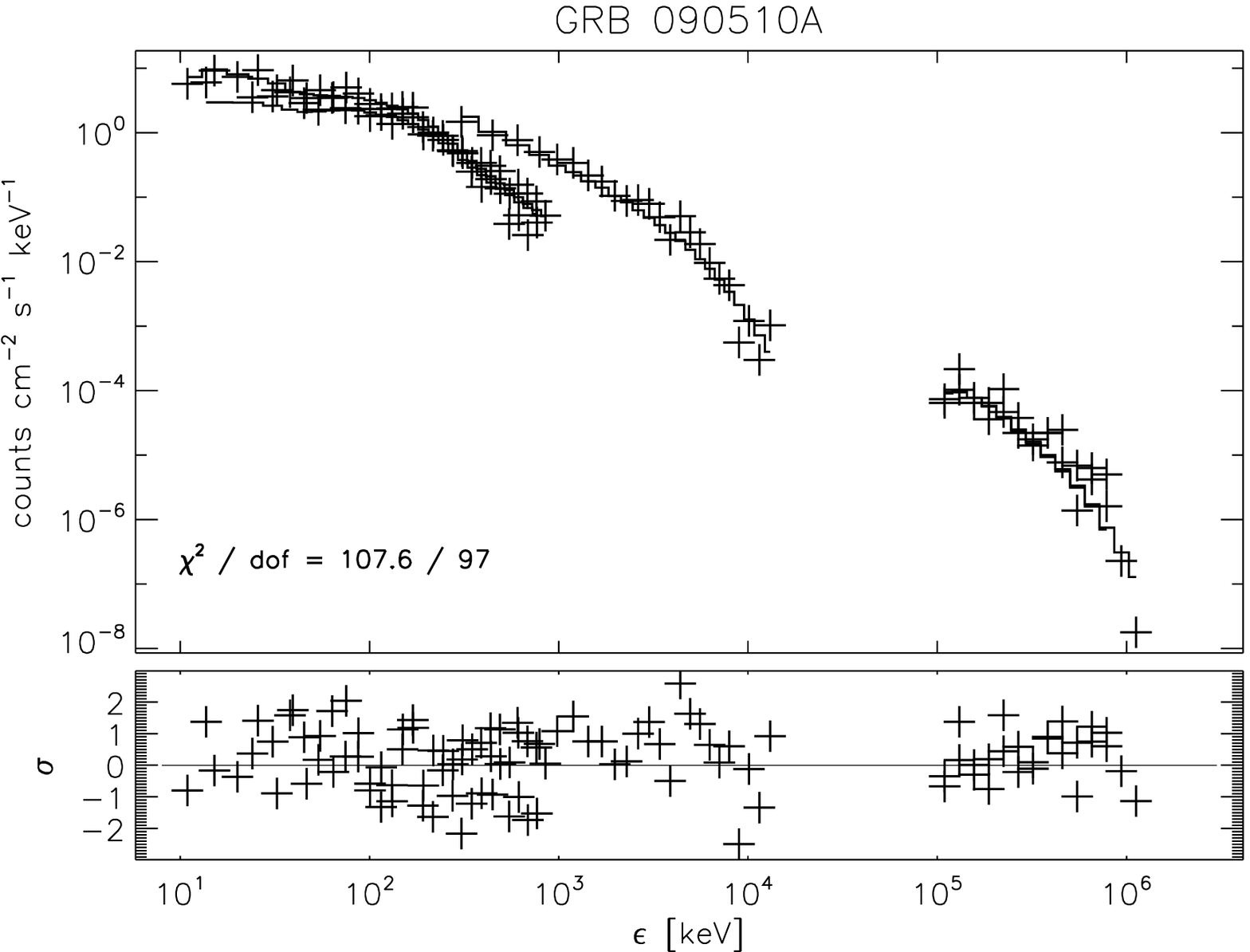}\\
\caption{GRB 090510 fit with a magnetic model ($\chi^2/$dof=107.6/97). In the
upper figure lines represent the model. The points with the error bars can be
thought of as measured values, though they are themselves model dependent (i.e.
for different models they are not the same). They are calculated from the model
convolved with the detectors' responses (step curve on lower figure) and their
relation to the observed counts (crosses on lower figure). For brevity we do
not show the plot with the convolved spectra for other bursts.	}
\label{fig:090510m} \end{center} \end{figure}

\begin{figure}[htbp]
\begin{center}
\includegraphics[width=.7\columnwidth]{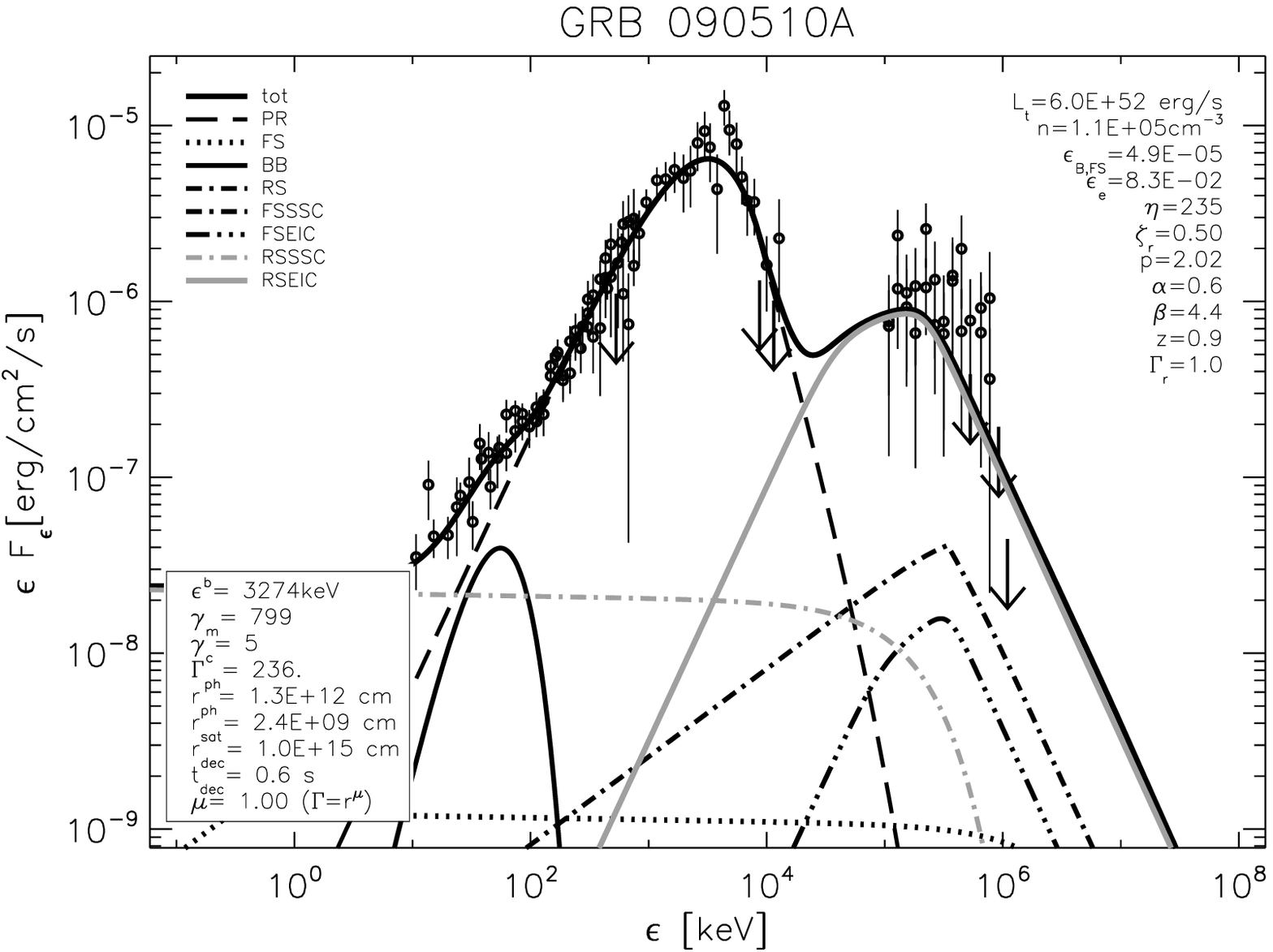}\\
\includegraphics[width=.7\columnwidth]{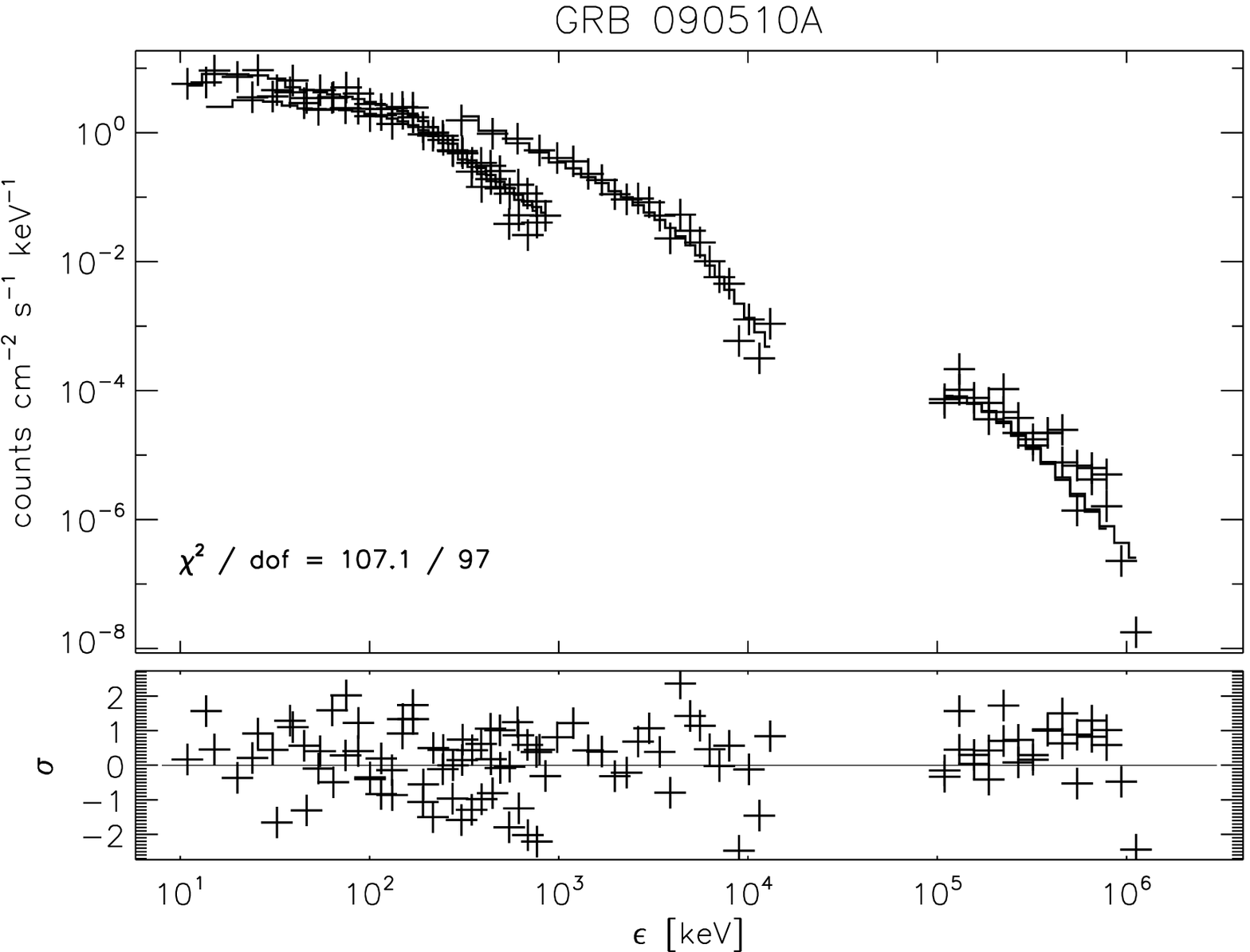}
\caption{GRB 090510 fit with a baryonic model ($\chi^2/$dof=107.1/97). Other
notations are the same as Fig. \ref{fig:090510m}  }
\label{fig:090510b} \end{center} \end{figure}

This is a short burst \citep{Abdo+09-090510} with  GeV emission.  According to
our modelling the time integrated spectrum can be described by a Band function
and the high energy photons are dominated by the RSEIC emission. The low-energy
excess is best modelled by a FS component in the magnetic case and by a
blackbody component in the baryonic case. See Figs. \ref{fig:090510m} and
\ref{fig:090510b}.

\subsection{GRB 080916C fit for time integrated spectrum}
\label{sec:080916c}
\begin{figure}[htbp]
\begin{center}
\includegraphics[width=.7\columnwidth]{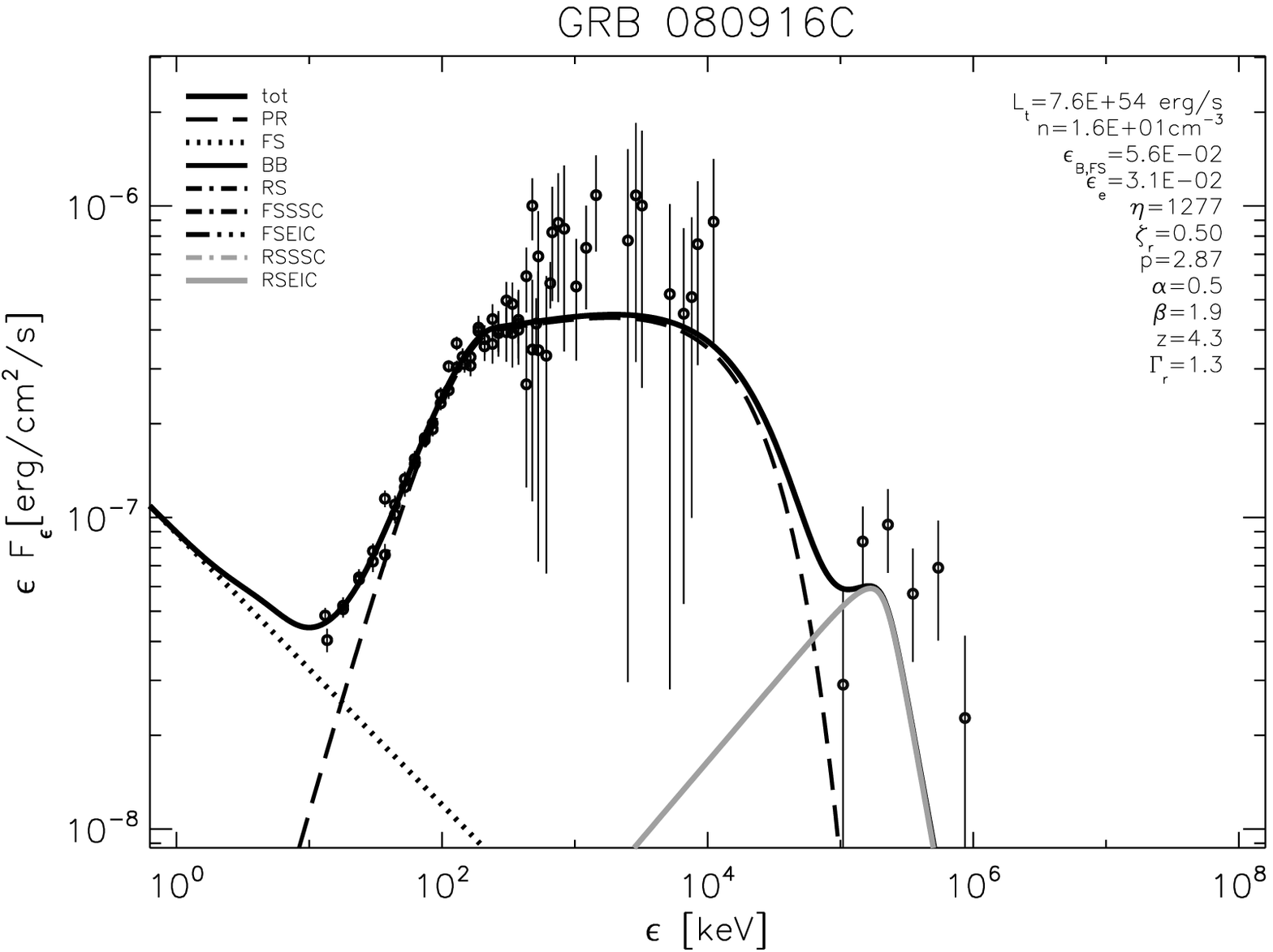}\\
\includegraphics[width=.7\columnwidth]{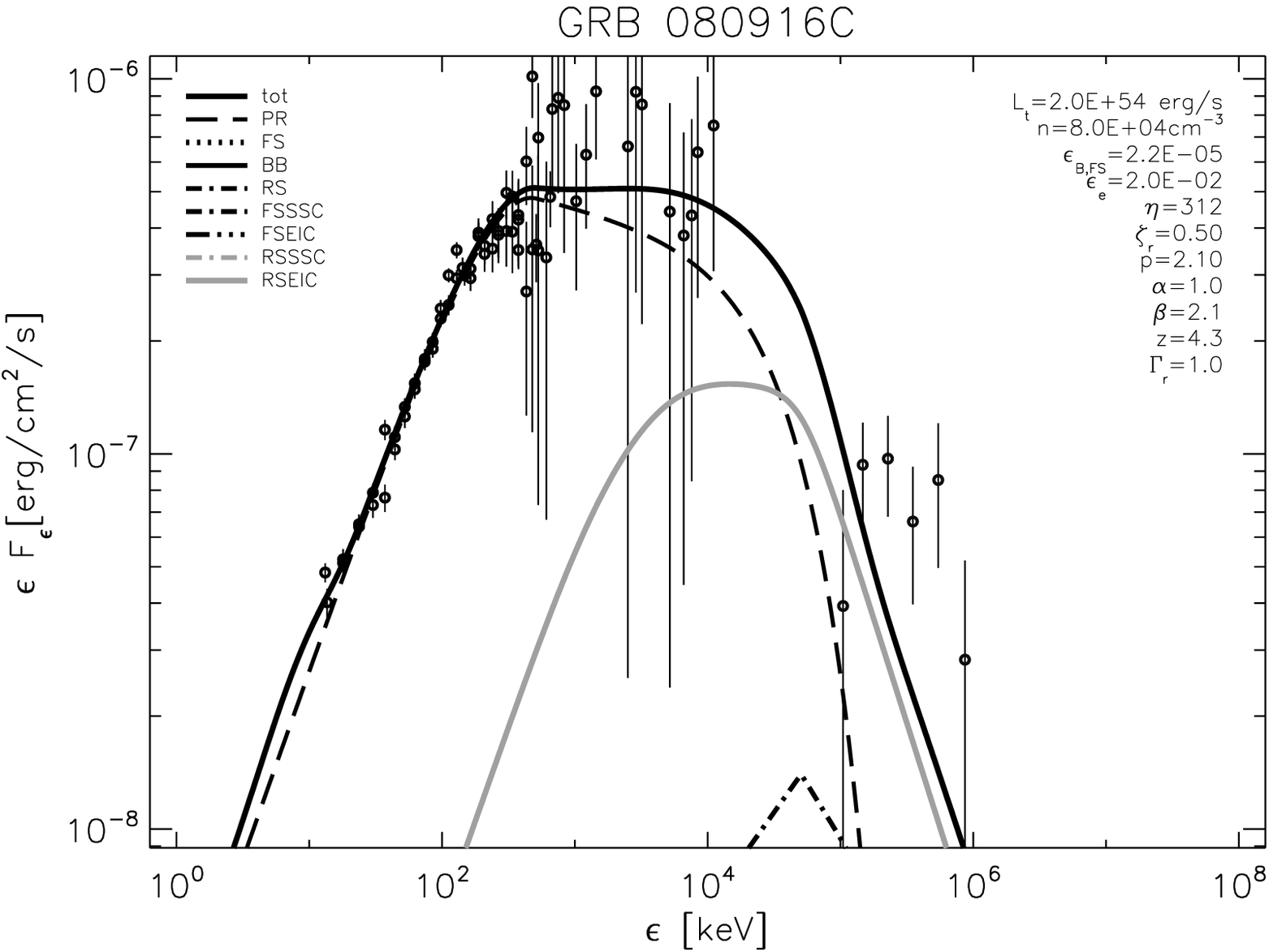}
\caption{
Fit for GRB 080916C, for the entire interval, with a magnetic model (above,
$\chi^2=113.6$, d.o.f.=$98$) and baryonic model (below, $\chi^2=129.7$,
d.o.f.=$98$). The figures show the models and the data points obtained by
scaling the measured count rates with respect to the  convolved models.
}\label{fig:080916cm} \end{center} \end{figure}

GRB 080916C is one of the most intensive bursts ever detected
\citep{Abdo+09-080916c}. Its spectrum is consistent with the Band function
throughout the Fermi energy range and in all the analyzed intervals (though
there was weak evidence of an extra component \citep{Abdo+09-080916c}).  The
{extremal} magnetic and baryonic fits are shown in Fig. \ref{fig:080916cm}.

\subsection{GRB 090926A fit for the time resolved spectrum}
\label{sec:090926}

\citet{Ackermann+11-090926} fitted the	spectrum of this burst	with a Band
function and an extra power-law with a high energy cutoff.  Using data from
\citet{Zhang+11-latgrb} the whole interval of the burst cannot be well fit with
a simple analytical model, suggesting strong spectral evolution. We thus use
the time resolved data for this GRB.
We divided the burst in 4 intervals (with limits at 0, 4, 7, 11.5 and 19 s with
respect to the trigger time). The first interval has no GeV range emission, and
we drop it from our analysis.

\begin{figure}[htbp]
\begin{center}
\includegraphics[width=.70\columnwidth]{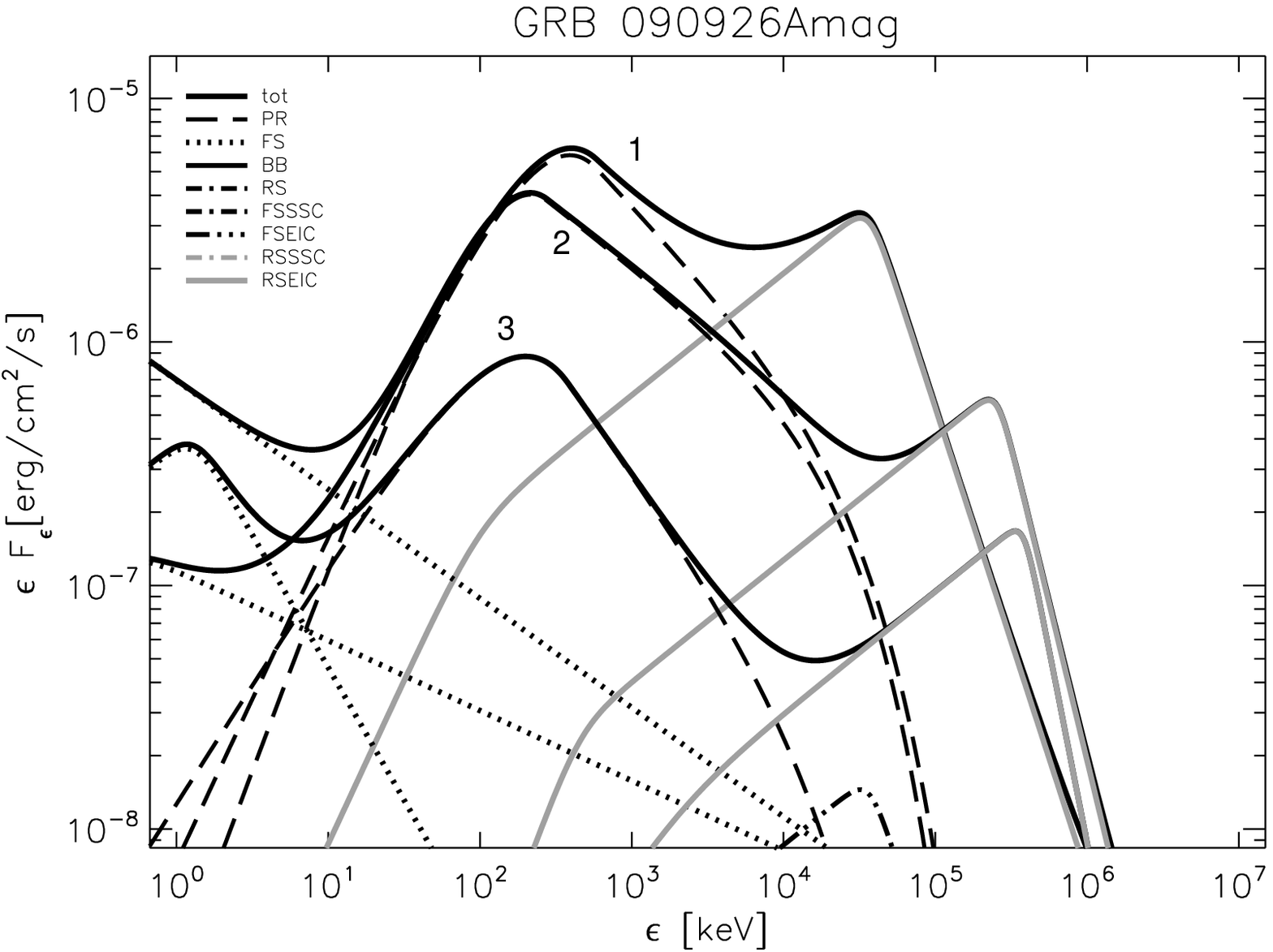}\\
\includegraphics[width=.70\columnwidth]{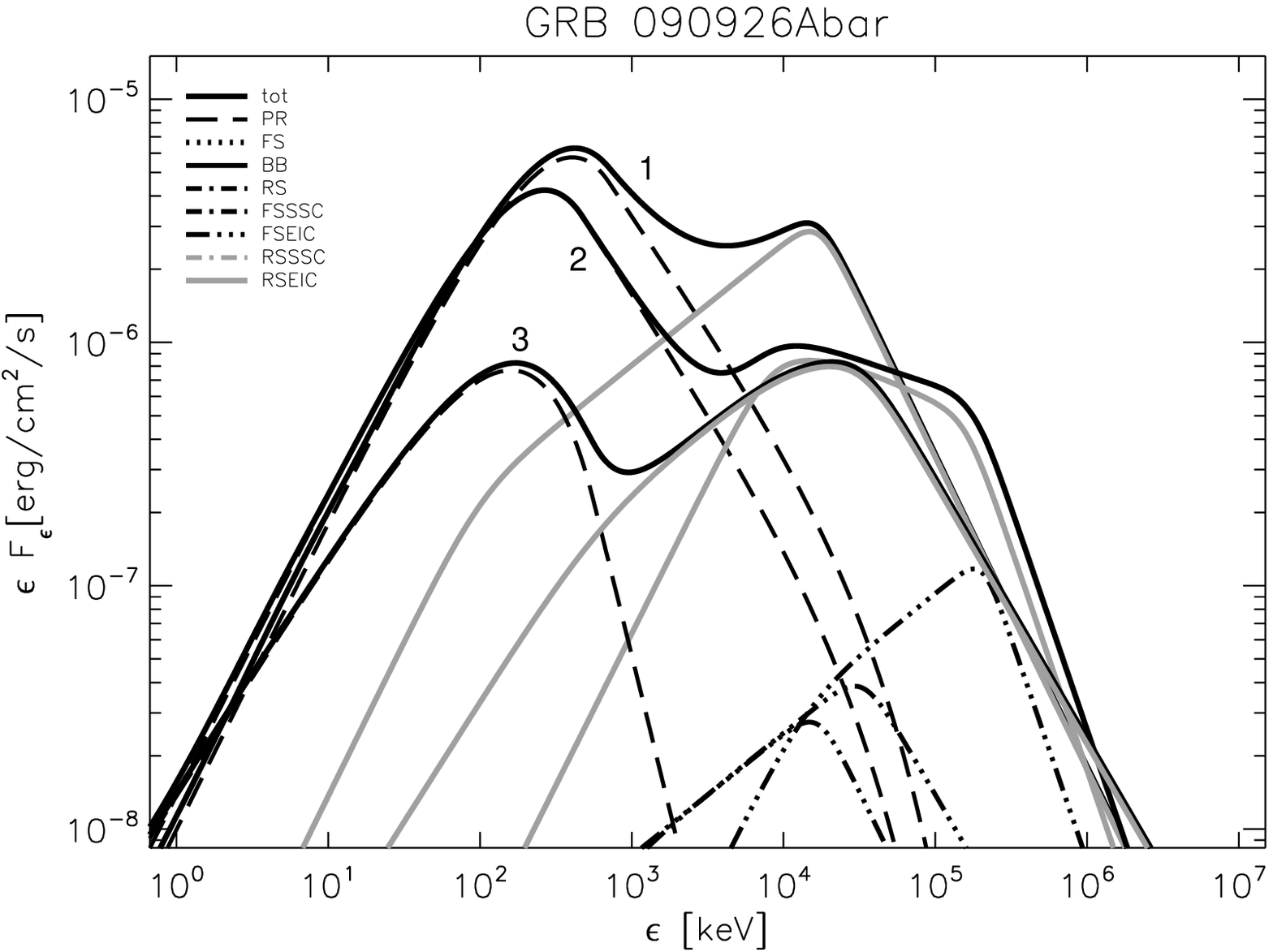}
\caption{GRB 090926A with three intervals fitted using the magnetic (top) and
baryonic (bottom) photosphere model. We have displayed only a small number of
components. The numbers indicate the intervals: 1: 4-7 s, 2: 7-11 and 3:
11.5-19 s. The 0-4 s period could not be fitted, because no LAT photons were
detected.  } \label{fig:090926Amultim} \end{center} \end{figure}

\begin{figure}[htbp]
\begin{center}
\includegraphics[width=.70\columnwidth]{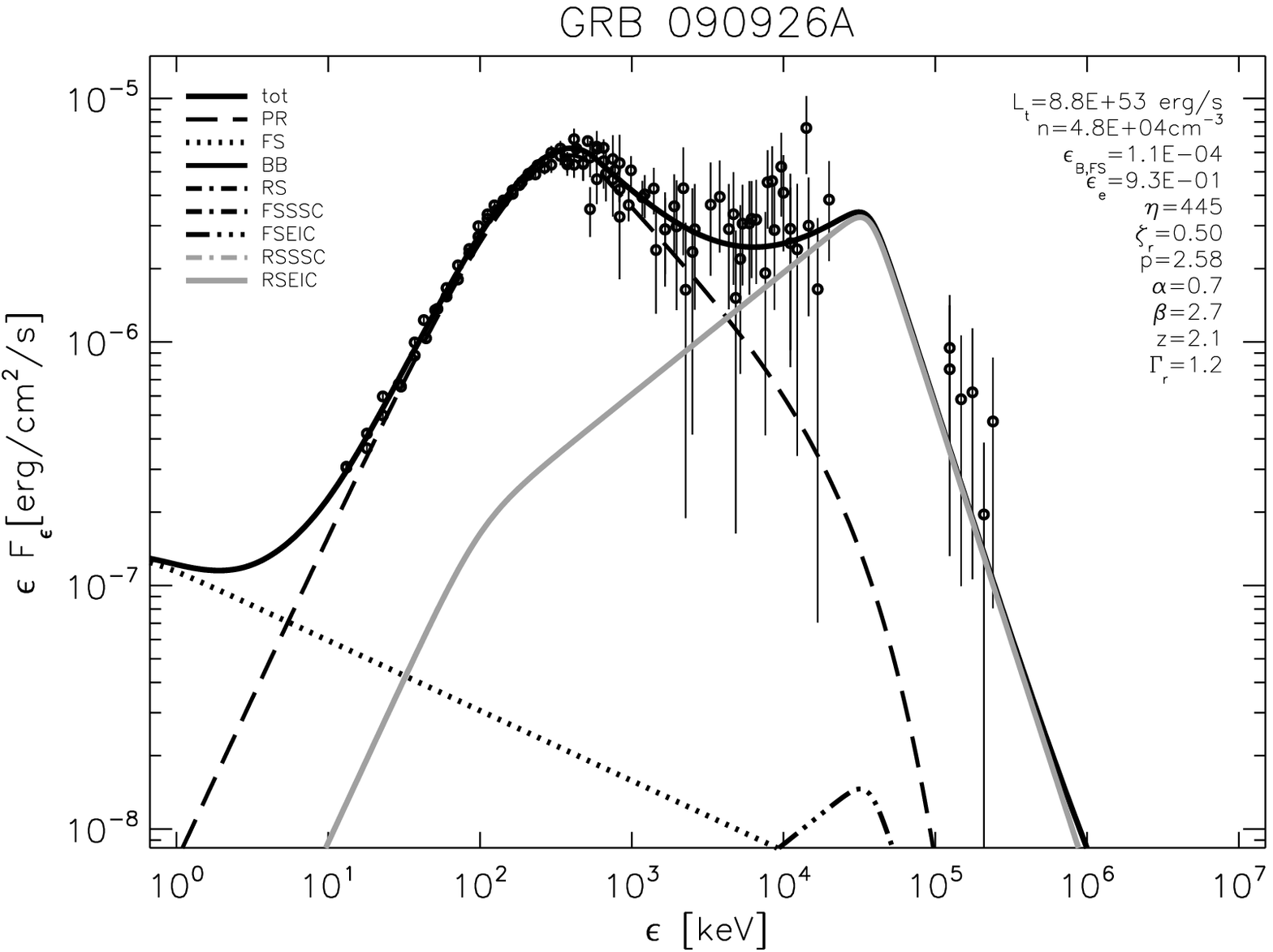}\\
\includegraphics[width=.70\columnwidth]{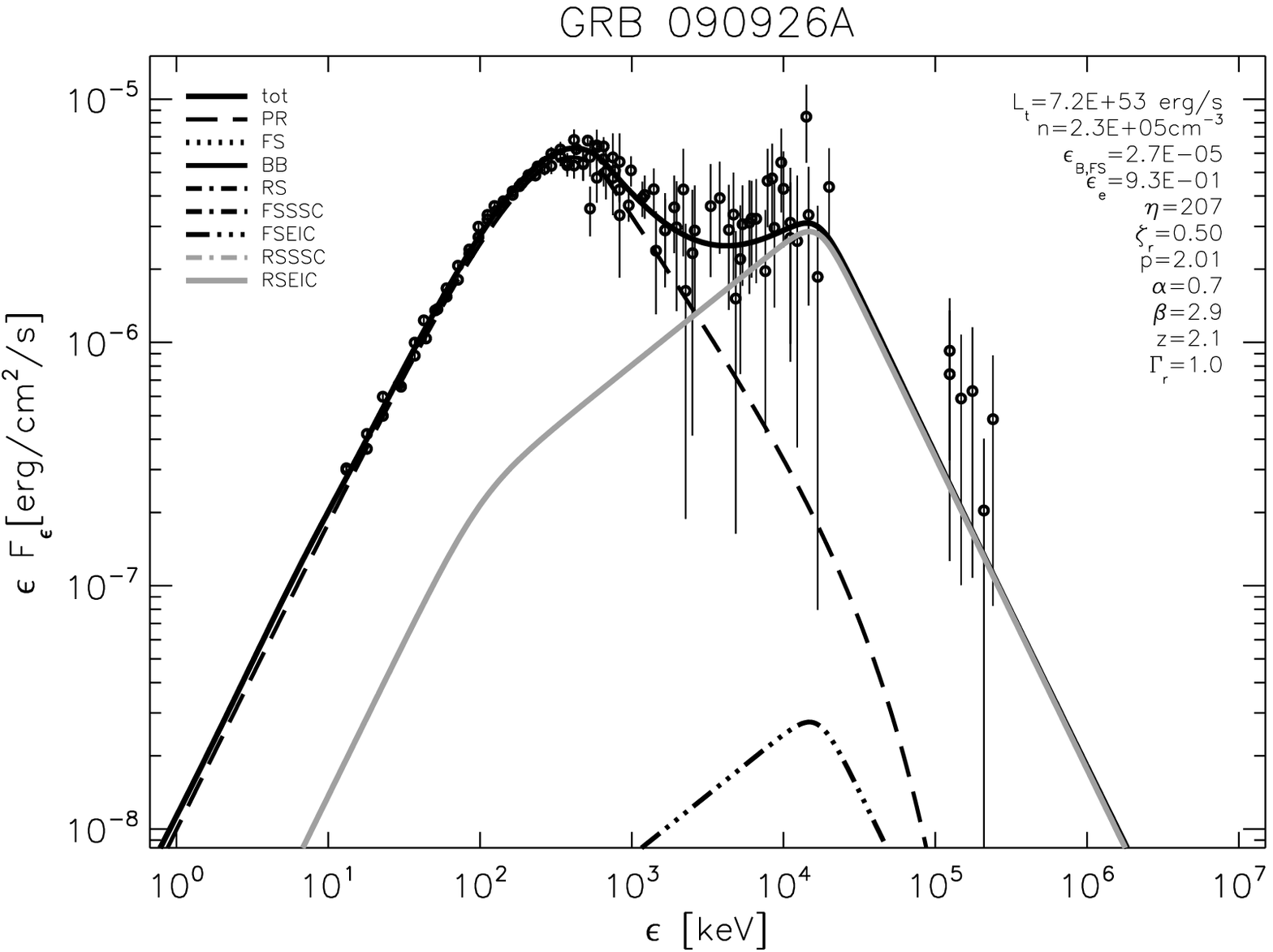}
\caption{Examples of magnetic (above, $\chi^2=182.9$, d.o.f.=$115$) and
baryonic (below, $\chi^2=184.5$, d.o.f.=$115$) models with the data overplotted
for GRB 090926A's  4-7 s interval (see also Fig. \ref{fig:090926Amultim}). }
\label{fig:090926A1} \end{center} \end{figure}

We found that both the magnetic and the baryonic photosphere model can give an
accurate description of the resolved data (see Fig. \ref{fig:090926Amultim} for
a superposition of magnetic and baryonic models and the models and data for the
first interval on Fig. \ref{fig:090926A1} ).

Reduced $\chi^2_r$ values are in the range $1.4-1.8$ for the magnetic model,
and $1.2-2.5$ in the baryonic case. Generally, $\epsilon_e$, $\epsilon_B$ and
terminal Lorentz factors ($\eta$) are systematically lower in the baryonic case,
the external density is systematically higher, while the other parameters are
roughly in the same range for both cases. We note as a general trend, the
magnetic model yielded better fits for the data.

\subsection{GRB 090902B fit for the time resolved spectra}
\label{sec:090902}

GRB 090902B \citep{Abdo+09-090902B} has a relatively sharp Band spectrum with
an underlying power law component. While a simple Planck function is too narrow
to fit the spectrum, a multicolor blackbody and an underlying power law can fit
the data equally well \citep{Ryde+10phot090902}. The fit for the entire
duration of the burst did not give an acceptable fit for neither of the models,
possibly due to the strong spectral evolution.	It has been shown that this
burst involves a strong spectral evolution \citep[e.g.][]{Zhang+11-latgrb}

\begin{figure}[htbp]
\begin{center}
\includegraphics[width=0.7\columnwidth]{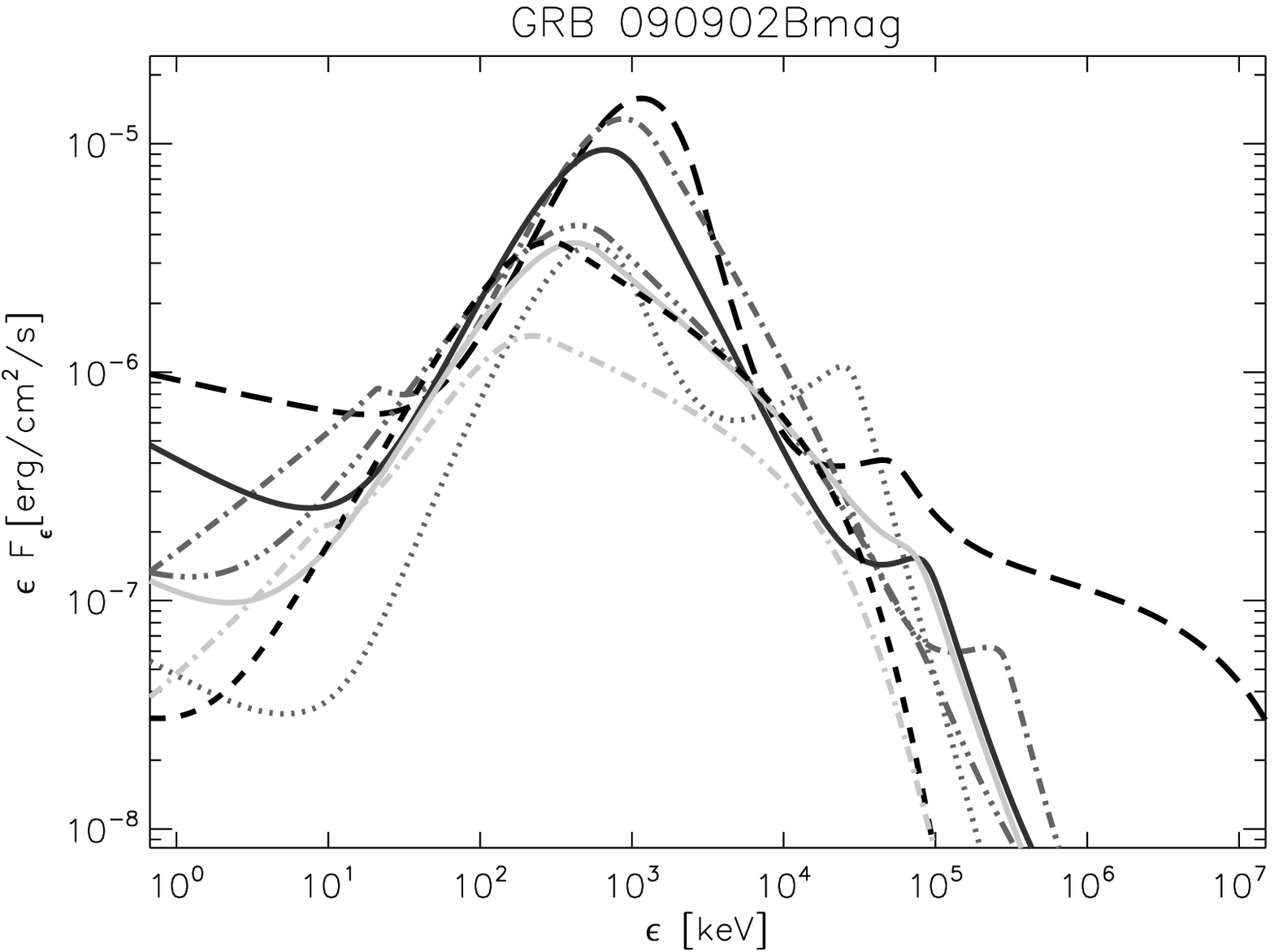}\\
\includegraphics[width=0.7\columnwidth]{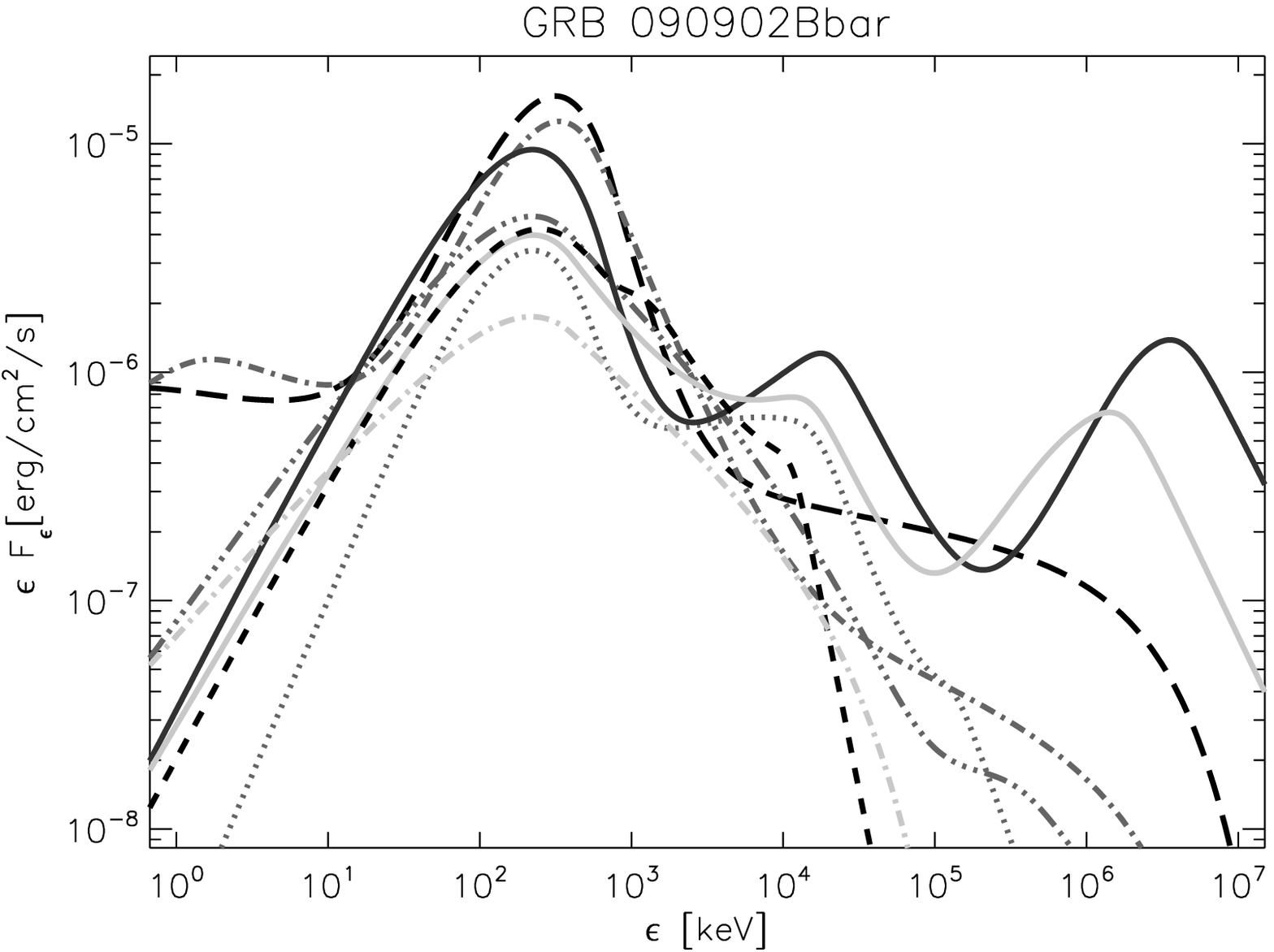}
\caption{GRB 090902B magnetic models (upper) and baryonic (lower). The time
intervals depicted: 0-5 s: gray-dotted, 7.3-9.7 s: black-long dashed, 9.7-12 s:
gray-dash-dotted, 12-14.3 s: gray-dash-dot-dot-dot, 14.3-16.7 s:
black-continuous, 16.7-19 s: gray-continuous, 19-21.3 s: black-dashed,
21.3-23.7 s: light gray-dash-dotted.  }\label{fig:090902B} \end{center}
\end{figure}

In order to account for spectral evolution, we have defined $10$ intervals
bracketed by the $0, 5, 7.3, 9.7, 12, 14.3, 16.7, 19, 21.3, 23.7$ and $26$ s
marks relative to trigger (see Fig. \ref{fig:090902B}).

\begin{figure}[htbp]
\begin{center}
\includegraphics[width=.70\columnwidth]{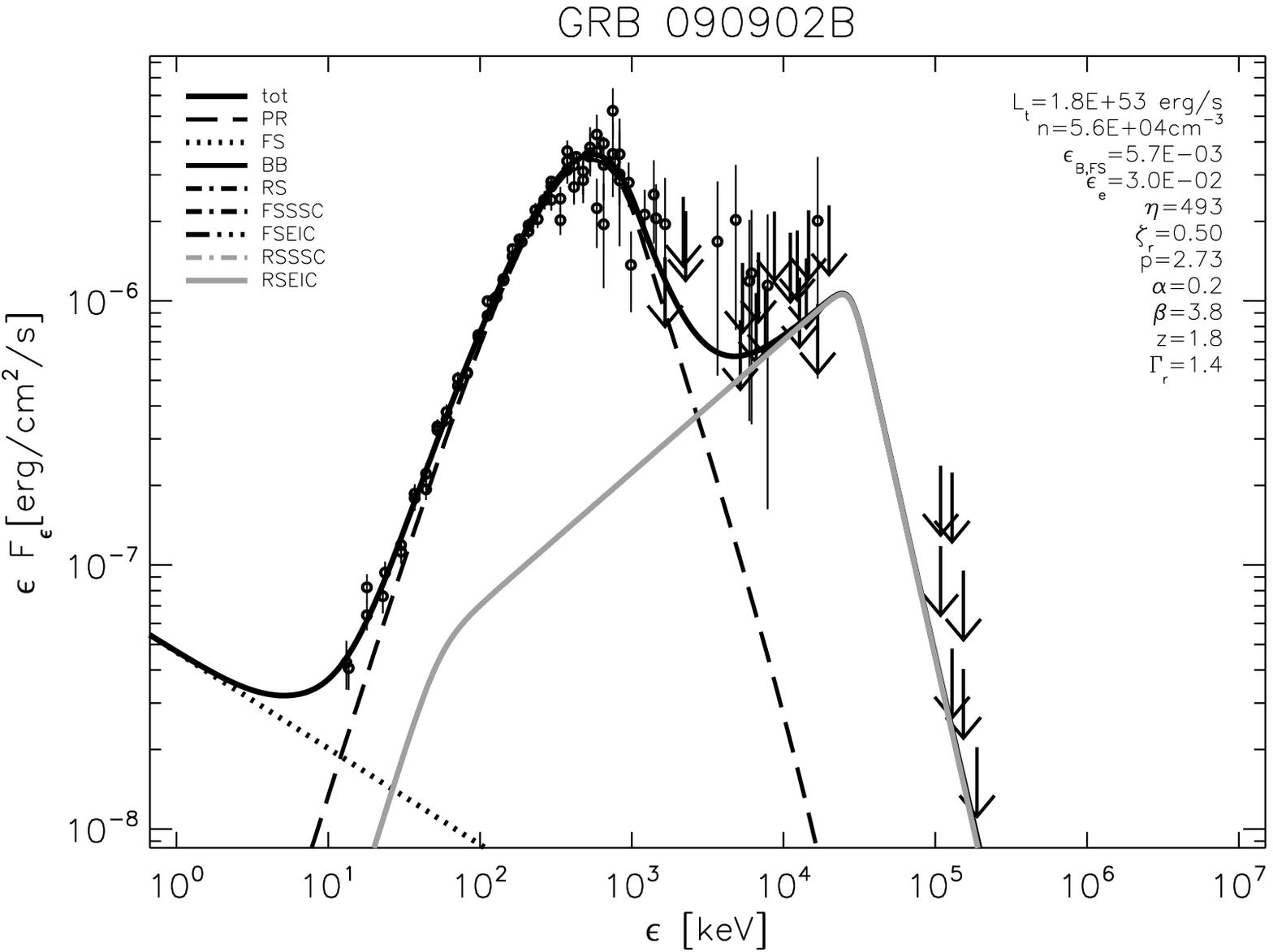}\\
\includegraphics[width=.70\columnwidth]{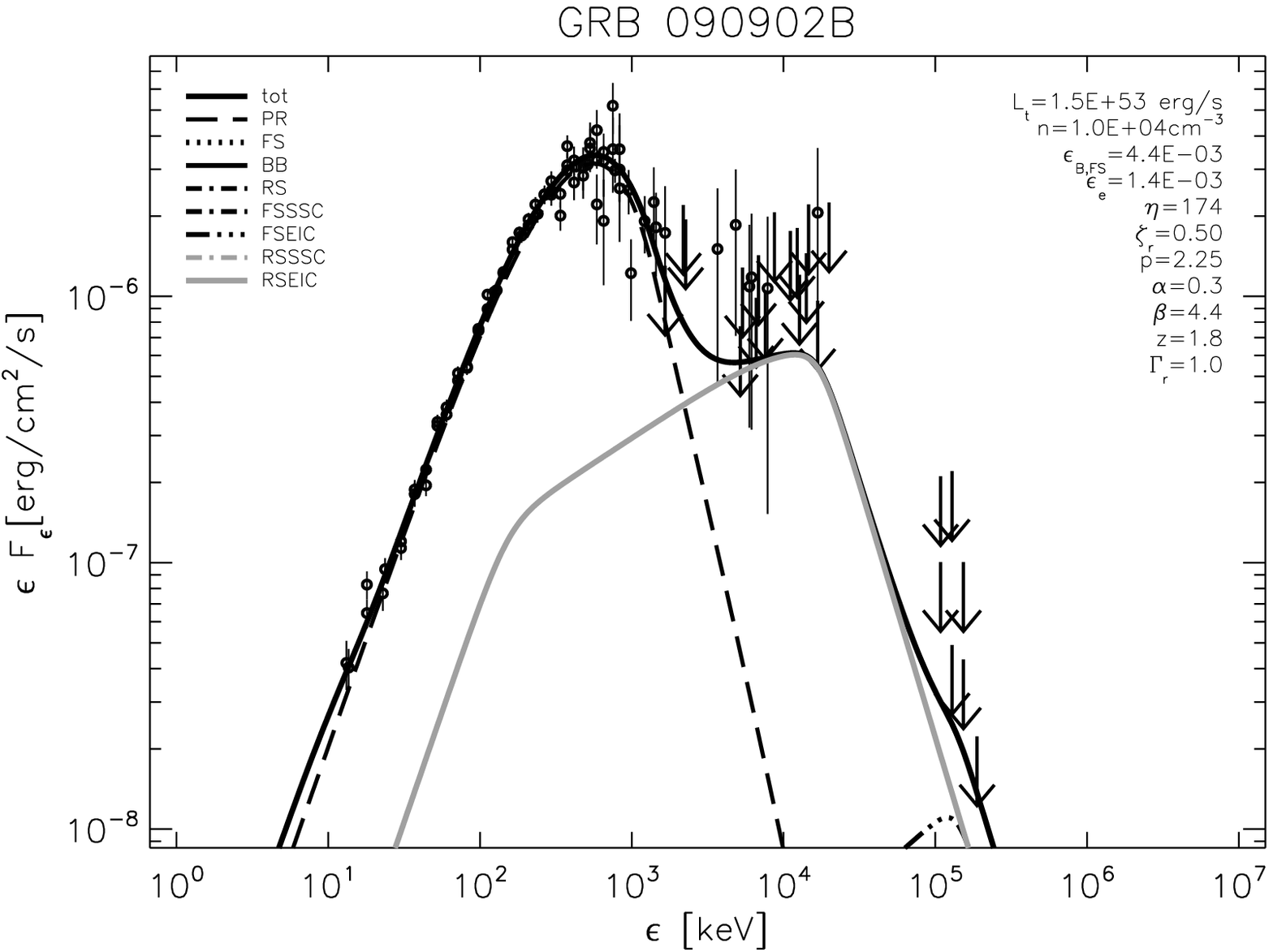}
\caption{Examples of magnetic (above, $\chi^2=102.3$, d.o.f.=$92$) and
baryonic (below, $\chi^2=96.7$, d.o.f.=$92$) models with the data overplotted
for GRB 090909B's  0-5 s interval (see also Fig. \ref{fig:090902B} -
gray-dotted curve). } \label{fig:090902B1} \end{center} \end{figure}

$\epsilon_e$ varies in the range $\approx 0.001-0.7$, and it is similar for
both models.  The terminal Lorentz factor is systematically higher in the
magnetic case and it is above $400$ in all intervals, while in the baryonic
case, it is below $250$. The external density, is systematically lower in the
magnetic case, while the magnetic parameter at the deceleration radius and the
semirelativistic shocks' Lorentz factor in the photosphere are systematically
higher in the magnetic case.	These parameters are all within the normal
ranges found previously from fitting afterglow and prompt emission data on a
wider sample of bursts.  Here, we do not find any obvious trend for an
evolution of the parameters throughout this burst. For an example fit for this
burst's first interval, see Fig. \ref{fig:090902B1}.

\section{Discussion}
\label{sec:disc}

{\it Possible relation between $\mu$ and observed trends.-}
In our fits we have purposefully avoided introducing any of the observed
phenomenological correlations claimed for GRBs, since these have errors and
biases which require extensive consideration and would complicate the analysis.
However, it is interesting to speculate on the possible values of $\mu$ in the
context of some of these phenomenological correlations. 

{In the $\eta>\eta_T$ case it seems simple considerations do not yield a
consistent value for $\mu$, and this needs to be examined further.  In the
other case ($\eta<\eta_T$), there is no $\mu$ dependence, but we only need to
assume one of the relations to obtain approximately the other one; e.g. if we
assume $\eta\propto L^{0.3}$ \citep{Lu+12corr}, we get $\ve_b\propto L^{0.4}$,
fairly close to the Yonetoku relation.}

Although we did not find any cases where the main MeV episode was provided by
the thermal component, {such cases have been discussed based based on more
elaborate radiation transfer models, e.g. by \citet{Peer+11multicolor,
Beloborodov12peak, Vurm+12therm} and others, typically for baryonic dynamics.
In our simplified treatment the thermal component was taken as a simple
blackbody function, instead of a broader multicolor blackbody as in, e.g.
\citep{Ryde+10phot090902}. Such multicolor fits including our other variables
would be much more cumbersome, but in principle, if we make a quick calculation
similar to that in the previous paragraph, using Eq.  \ref{eq:temp} instead of
Eq. \ref{eq:ebreak}, this would imply $\mu\approx 0.5$.  }

It is interesting to note that for the magnetically dominated model
($\eta>\eta_T$ and $\mu=1/3$), the peak energy does not depend on either the
luminosity and the terminal Lorentz factor, and has a value of $\sim 300 \kev$,
corresponding to the clustering, observed by e.g. BATSE
\citep{Kaneko+06BATSEcat}.  There is a weak dependence on $r_0$ and redshift.

{\it Comparison of the baryonic photosphere model with the magnetic model.-}
The main difference between the magnetic and the baryonic model in terms of the
physical parameters involved in the GRBs is the initial dependence of the
Lorentz factor on the radius (see equ. \ref{eq:accel}).

The main manifestation of this difference in terms of the observed spectrum is
the strength of the thermal component from the photosphere. Because of the
different dynamics, the magnetic model results in a sub-dominant thermal
component, while in the case of the baryonic dynamics the black body component
can have a more prominent contribution.

The main goal of this article has been to present a generalized formalism for
models with an arbitrary acceleration of the GRB ejecta, valid for magnetic and baryonic
dynamics. By means of an analysis of the brightest bursts, and properly accounting
for the instrumental effects, we have shown that both an {extremal}  magnetic and a
baryonic model can provide adequate fits to the spectrum. Although in some of the cases
the magnetic model provides a marginally better fit to the data, the difference is
not significant enough to decide in favor of either extremal model.


{\it Constraints on the parameters.-}
As we can see from Table 1. we can constrain some of the physical
parameters of the models, except for the magnetic parameter at the deceleration
radius, and in some cases the interstellar density is also difficult to constrain.

We found that the burst spectra are consistent with both the magnetic and the
baryonic models presented here.  A possible discrimination would require further
theoretical considerations, or else other observational considerations; e.g. the
length of the deceleration time should be the same order as the observed delay of the GeV
emission. Another discriminant would be how well a model can reproduce the observed
correlations between the rest-frame peak energy and e.g. luminosity would.  As we
indicated above, the extreme magnetic photosphere model appears to be  better in
accommodating these constraints.

One of the main differences between the two models is the systematically higher
value of the coasting Lorentz factor (LF) in the magnetic model. This can be understood
as follows: by applying two models to the same data the photospheric LF is the
best constrained by the observations. While in the baryonic case this is just
the terminal LF, in the magnetic case the ejecta is still accelerating at the
photosphere, having a larger coasting LF.


In GRB 090510 \citet{Abdo+09-090510} noted a correlation between the keV range
radiation and the GeV radiation (but no correlation with the MeV radiation).
Based on their spectrum, the additional power-law would be the responsible for
explaining this correlation.
In the framework of this generalized photospheric model, such a correlation can be
explained if these two components have a common source: e.g.  the keV emission
is ascribed to the forward shock (FS) synchrotron and the GeV to the forward shock
synchrotron-self Compton (FS-SSC); or if a reverse shock is present (as proposed in
this work), the keV radiation can be due to the reverse shock synchrotron-self
Compton (RS-SSC) and the GeV is produced by the reverse shock external inverse
Compton (RS-EIC).

{\it Conclusions.-}
In summary, we have shown that a simple dissipative photosphere model of the
prompt emission of GRB based on synchrotron radiation can fit well the joint
Fermi GBM and LAT data for the four brightest bursts with sufficiently detailed
information. We have introduced a general formalism for treating the jet
acceleration dynamics ranging between an extreme magnetically dominated to a
baryon dominated dynamics regime. Both types of dynamics give acceptable fits
to the data on these four bursts. Additional considerations involving statistical
trends over a much larger sample of bursts might favor slightly the magnetic
models, but firm conclusions would require detailed fits over much larger samples
than undertaken here.

We acknowledge partial support from NASA NNX09AL40G and OTKA K077795 (PV, PM)
and NASA SAO SV4-74018 and NASA SAO GO1-12102X (BBZ). We thank Kazumi Kashiyama, Yi-Zhong Fan and the anonymous referee for comments and suggestions.


\end{document}